\documentstyle[12pt]{article}
\begin{document}
\setlength{\topmargin}{-1cm}
\setlength{\oddsidemargin}{0cm}
\setlength{\evensidemargin}{0cm}
\title{
\begin{flushright}
{\large \bf CERN-TH/95-288}
\end{flushright}
\vspace{1cm}
 {\Large\bf LIGHT COLOUR-TRIPLET HIGGS IS COMPATIBLE WITH PROTON STABILITY:\\
an alternative approach to the doublet-triplet splitting problem }}

\author{{\bf Gia  Dvali}\thanks{E-mail:
dvali@surya11.cern.ch }\\ CERN, CH - 1211 Geneva 23, Switzerland\\}

\date{ }
\maketitle

\begin{abstract}
 It is usually assumed that the proton stability requires the
coloured triplet partner of the electroweak
Higgs doublet to be superheavy (with a mass $\sim M_{GUT}$).
We show that this is a very model-dependent statement and the
colour triplet can be as light as the weak doublet without
leading to the proton decay problem. This implies an alternative
approach to the doublet--triplet splitting problem: instead of
using the mass difference the splitting can occur between the doublet and
triplet Yukawa coupling constants so that the
light Higgs triplet can appear decoupled from the
quarks and leptons and can not lead to the proton decay.
In this scenario the GUT symmetry breaking automatically induces an
extremely strong suppression
$\sim M_W/M_{GUT}$ of the coloured Higgs effective Yukawa coupling;
this happens
without any fine--tuning, just because of the Clebsch factors. Conceptual
differences of the above picture are: (1) an essentially stable proton:
both $d = 5$ and $d = 6$ proton decay mediating operators are suppressed
by the same factors $\sim (M_w/M_{GUT})^2$; (2) the possibility of
solving the $\mu$ problem by the light gauge singlet field
(this fact would lead to the destabilization of the hierarchy in the
standard case); (3) the existence of the long--lived, light,
coloured and charged supermultiplet in the $100$ GeV -- $TeV$
mass region, which can be the subject of an experimental search.
 We construct two explicit $SO(10)$ examples with the above properties, with
superpotentials most general under the symmetries. In both models, the
Higgs sector automatically delivers certain light states
which in combination with the coloured triplet form a complete $SU(5)$
multiplet, so that the unification of couplings is unaltered.
\end{abstract}

\newpage

\subsection*{1. Introduction}

 Perhaps the most difficult problem of the supersymmetric
grand unified theories (GUTs) is the `doublet--triplet splitting' problem.
The heart of the problem has to do with the fact that in the GUT
context the Higgs doublets of the minimal supersymmetric standard model
(MSSM) $H, \bar H$ $inevitably$ get accompanied by their coloured triplet
partners $T,\bar T$. GUT symmetry ($G$) forces the coloured triplet to be
coupled to the quark and lepton superfields ($Q,u_c,d_c,L,e_c,\nu_c$)
by the Yukawa coupling constant $Y_{T(\bar T)}$, which is  equal
(in the unbroken $G$ limit) to the one of the
doublet $Y_{H(\bar H)}$. In such a
situation the coloured triplet exchange can lead to an
unacceptably rapid proton decay unless $T, \bar T$ are superheavy.
This heaviness is certainly possible  since, unless forbidden by some
symmetry, the triplet can get large ($\sim M_{GUT}$) mass from the
couplings with the vacuum expectation values (VEVs) that  break $G$.
What is much, much more difficult, however, is to protect the doublet
partners from getting the same order mass. This is the famous doublet--
triplet splitting problem. Most of the attempts reported in the literature
deal with this difficulty. In certain approaches
such as the `missing partner' or the `missing VEV' \cite{mp}, the
doublet appears light
for the group theoretical reasons or because of the VEV structure.
In the `pseudo--Goldstone picture'\cite{pp} its mass is protected
by the Goldstone theorem
and is fully controlled by the scale of SUSY breaking. The common
feature of these approaches is that they try (though in different ways)
to make the colour triplet very heavy in order to suppress the proton
decay.

Certainly there is a loophole that may avoid such an approach:
there is no need for the heavy triplet if its effective Yukawa coupling
constant $Y_T$ is suppressed by many orders of magnitude with respect
to the one of the doublet $Y_H$. Say, if $Y_T/Y_H \sim M_W/M_{GUT}$,
such a triplet can never lead to an observable proton decay even if
its mass is comparable with the weak scale $M_W \sim $100 GeV --TeV.
As suggested in \cite{dv},
such a situation can occur without any fine--tuning if the matter fermion
masses are originated from the effective high--dimensional operators
induced by the physics at $M_G$. To be more explicit consider the following
$SO(10)$ invariant operator \cite{dv}
\begin{equation}
    {Y_{\alpha, \beta} \over M}   10_i 45_{ik}16^{\alpha}\gamma_k 16^{\beta},
\end{equation}
where $16^\alpha \alpha =1,2,3$ are three families of the matter fermions,
$10_i$ ($i=1,..., 10$) is the  multiplet in which reside the
$H,\bar H \in 10_i (i=7,..., 10)$
and $T, \bar T \in 10_i (i= 1,..., 6)$  states and $45$ is the
GUT Higgs in the adjoint representation of $SO(10)$. We have
written $SO(10)$ tensor indices
($i,k$) explicitly, since the way of their construction is important for us
and $\gamma_i$ are the matrices of the $SO(10)$ Clifford
algebra. $M$  is a certain regulator scale $\sim M_{GUT}$. Coupling $(1)$
has to be understood as an effective operator obtained through the
integrating out of some heavy states at $M_{GUT}$. Below we show explicitly
how the above structure can result automatically from the
tree--level exchanges \cite{bd}
of heavy `scalar' or `fermionic' superfields with purely $renormalizable$
interactions. Before doing this, let us simply assume for a moment
that coupling $(1)$ exists due to whatever reason and see its role in the
D--T splitting problem. For this we require the $45$-plet Higgs to have the
VEV of the form
\begin{equation}
\langle 45_{ik}\rangle = diag[0,0,0,A,A]\otimes \epsilon
\end{equation}
where $A \sim M_{GUT}$ and each element is assumed to be proportional
to the $2 \times 2$ antisymmetric matrix $\epsilon$. This  VEV breaks
the $G_{L,R} = SU(2)_L\otimes SU(2)_R \otimes SU(4)$ subgroup of $SO(10)$
down to $SU(2)_L\otimes U(1)_R \otimes SU(4)$ and is thus oriented along the
$T_R^3$ generator of $SU(2)_R$.  In combination with the other VEVs,
say the $16$-plet with non-zero SU(5)-singlet VEV ($\nu_c$), it leads
to the desired breaking $SO(10) \rightarrow G_W =SU(3)_c\otimes SU(2)_L
\otimes U(1)_Y$. Now, inserting (2) in (1), we do not need much effort
to be convinced that the effective Yukawa coupling constants
of the Higgs doublets are $Y_{\alpha,\beta}^H = Y_{\alpha,\beta} {A \over M}$,
whereas the triplets  have no couplings at all! Such a decoupled triplet
cannot lead to the unacceptable proton decay, even if it is as light
as its doublet partner. Thus, there is no need to couple the $10$-plet to
the GUT Higgses and the unnatural fine tuning can be avoided. Of course,
in this approach the standard coupling $16^{\alpha}16^{\beta}10$ must
be forbidden by some symmetry. Therefore, we see that the $10$-plet must
transform under a certain symmetry $G_{10}$ such that: (1) it prevents
the $10$ from coupling with the GUT Higgses; (2) it allows $10$-plet
to couple with $16^{\alpha}$ $only$ in combination with $45$-plet . Below
we will show that the solution of the $\mu$ problem fixes
$G_{10} = Z_2\otimes Z_3$.

 What are the motivations for the above approach? First we
see at least three model-independent interesting consequences:

(1) It automatically solves the problem of the coloured-Higgsino-mediated
proton decay via $d=5$ operators \cite{d5}: In contrast with a standard case
the suppression occurs not because of the large mass, but of
the small coupling. As a result both $d=5$ and $d=6$
operators are suppressed by factors at least
$\sim (M_w/M_{GUT})^2$ and the proton is practically stable;

(2) In contrast with the standard GUTs, the $\mu$ problem can be solved
through the coupling of the Higgs doublet to the light
gauge singlet field. In the
normal case this would lead to a destabilization of the hierarchy through
the well-known `tadpole' diagram \cite{tp}. For the obvious reasons,
no such diagram
exists in the present case.

(3) Prediction of the existence of long-lived $T,\bar T$ supermultiplets
in the $100$ GeV - TeV mass region, which can be the subject of
an experimental
search.

 Last but not least, this approach is an alternative to the existing
schemes of the D--T splitting problem and is certainly worth studying,
especially since there exists no fully satisfactory solution
at present. In this paper we construct the two realistic $SO(10)$
GUTs which explicitly realize the above scenario. Both models have
superpotentials most general under symmetries and do not require any
fine-tuning or unnaturally small parameters.

\subsection*{2. The mechanism}

In this section we study the origin of the operator (1) in somewhat
more detail. As was mentioned, it can be generated by the
exchange of the intermediate `fermionic' or `scalar' superfields.
The reader should not be confused with these names, since they simply
refer to the components propagating in the internal line of the
tree (super)diagram
once we fix the external legs to be fermions for
$16^{\alpha}$ and scalars for $10$ and $45$. The `fermionic' exchange
was already studied in \cite{dv}, but we will consider it in detail,
since it is an essential ingredient for our models.

\subsubsection*{2.1 Fermionic exchange}

In order to generate (1) through the heavy-fermion exchange we
introduce the three pairs of superfields  $144_{\alpha},
\overline{144}_{\alpha} (\alpha = 1,2,3)$, where their representation
content under $SO(10)$ is indicated explicitly. The relevant
piece of the superpotential has the form
\begin{equation}
 W_{Yukawa} = g_{\alpha}^{\beta}16^{\alpha}\overline{144}_{\beta}45
 + M^{\alpha\beta}144_{\alpha}\overline{144}_{\beta} +
g_{\alpha}^{'\beta}16^{\alpha}144_{\beta}10.
\end{equation}
This superpotential is invariant under the discrete $Z_2\otimes Z_3$
symmetry acting in the following way:

under $Z_2$: only $10,45, 144$ and $\overline{144}$ change sign;

under $Z_3$: $10 \rightarrow e^{i2\theta} 10$;
$(16, 144) \rightarrow e^{-i\theta} (16, 144)$, and
$ \overline{144} \rightarrow
e^{i\theta}\bar {144}$.

In fact, any $\theta$ leaves $W_{Yukawa}$ invariant, but as shown below
only $\theta= 2\pi/3$ is allowed by the solution of the $\mu$ problem.
This $Z_3$ symmetry ensures that $10$ cannot couple bilinearly
 with any of the GUT Higgses in the superpotential and both doublet
and triplet are light. For energies much below $M$, this superpotential
is effectively equivalent to the operator (1), due to a
structure of $144$ representation. In order to be sure that the above
structure can indeed suppress the proton decay, we have to find the effective
couplings of $T,\bar T$ with the quarks and leptons. The strategy is
straightforward: (1) insert the VEV of $45$ in (3); (2) find the light
superpositions (express initial states through the final mass eigenstates);
(3) insert the answer in the operator
$g_{\alpha}^{'\beta}16^{\alpha}144_{\beta}10$ and select only the
potentially dangerous couplings $T(\bar T)$ -- light -- light. The
$G_{LR}$ decompositions of some $SO(10)$ representations
are very useful \cite{sl}:
\begin{eqnarray}
10 &=& (2,2,1) + (1,1,6)\nonumber\\
16 &=& (2,1,4) + (1,2,\bar 4)\nonumber\\
144 &=& (2,1,4) + (1,2,\bar 4) + (3,2,\bar 4) + (2,3,4) + (2,1,20) +
(1,2,\overline{20})\nonumber\\
45 &=& (1,3,1) + \dots
\end{eqnarray}
The $T,\bar T$ triplets live in $(1,1,6)_{10}$. Thus,
its $G_{LR}$-invariant couplings with $16$ and $144$ are
\begin{eqnarray}
&& g_{\alpha}^{'\beta} (1,1,6)_{10}
\bigl[ (2,1,4)^{\alpha}_{16} (2.1.4)_{\beta}^{144} +
(2,1,4)^{\alpha}_{16} (2.1.20)_{\beta}^{144}\nonumber\\
&& + (1,2,\bar 4)^{\alpha}_{16} (1,2,\bar {20})_{\beta}^{144}
+ (1,2,\bar 4)^{\alpha}_{16} (1,2,\bar 4)_{\beta}^{144}\bigr]
\end{eqnarray}
We now have to find the light admixture in these fragments. First of all,
we immediately notice that there is no state in $16$ to which
$(2,1,4)^{144}, (2,1,20)^{144}$ and $(1,2,\overline{20})^{144}$
can mix via $(1,3,1)_{45}$ VEV (the would existing states
had to transform as $(2,3,4), (2,3,20), (1,2,\overline{20})$ and
$(1,4,\bar {20})$, respectively).
Thus, $(2,1,4)^{144}, (2,1,20)^{144}$ and
$(1,2,\overline{20})^{144}$ are the purely
heavy states and the only potentially dangerous coupling in (4) is
the last one. Before discussing its strength, note that this
coupling involves only the $SU(2)_R$-doublet quarks and leptons. Thus,
the only possible $T$-light-light couplings are:
\begin{equation}
\bar T u_cd_c + T u_ce_c
\end{equation}
Even if not suppressed, this couplings can lead to the proton decay
only if there is a $T\bar T$ mass insertion somewhere. This is not
necessary in general, since $T,\bar T$ can get masses from mixing
with the other states. The latter can be an interesting possibility
$per~se$, but it is not
necessary in our case, since the above couplings
can be naturally absent. To see this, notice that
since mixing goes through the VEV $(1,3,1)$ the
resulting coupling in terms of the light mass eigenstates has the form:
\begin{equation}
 (1,1,6)_{10}<(1,3,1)_{45}>
(1,2,\bar 4)^{\alpha}_{light} (1,2,\bar 4)_{\beta}^{light}.
 \end{equation}
This coupling is antisymmetric in $SU(4)$ indices and symmetric in
$SU(2)_R$ ones. So it will automatically vanish if the Yukawa coupling
constants are symmetric in $\alpha,\beta$. This can be ensured by some
flavour symmetry, which sooner or later probably has to be invented
anyway in order to solve the fermion-mass problem.
As far as we are not
going to address this issue here, we will give just one possible
example of such a flavour symmetry: $SU(3)_f$, under which $16^{\alpha}$
and $144_{\alpha}, \overline{144}_{\alpha}$ are antitriplets and triplets,
respectively, and which is broken only by the Higgses in the
symmetric representation ($6$-plets). In such a case
$g_{\alpha}^{\beta} = g\delta_{\alpha}^{\beta},~
g_{\alpha}^{'\beta} = g'\delta_{\alpha}^{\beta}$ and
$M_{\alpha\beta}$ has to be understood as a VEV of the symmetric
representation of $SU(3)_f$.

\subsubsection*{2.2 Scalar exchange}

In order to generate eq(1) via the heavy scalar exchange
let us (instead of
$144,\overline {144}$) introduce a pair of $10',10''$-plets. The superpotential
of the Yukawa sector now becomes:
\begin{equation}
W_{Yukawa} = g_{\alpha\beta} 16^{\alpha}16^{\beta} 10''+
g'4510'10 + M10'10''.
\end{equation}

This superpotential is also invariant under a $Z_2\otimes Z_3$ symmetry
such that: under $Z_2$, $45$ and $10$ change sign, and under $Z_3$,
\begin{equation}
16 \rightarrow e^{-i\theta} 16,~~~
(10,10'') \rightarrow e^{i2\theta} (10,10''),~~~
10' \rightarrow e^{-i2\theta} 10'.
\end{equation}
After insertion of the $45$ VEV the mass matrices of doublets and
triplets become
\begin{equation}
(g'AH + MH'')\bar H' + (- g'A \bar H + M\bar H'') H'
+  M(T'\bar T'' + T''\bar T').
\end{equation}

We see that there is an admixture $\sim g'A/((g'A)^2 + M^2)^{1/2}$
of the light doublet in the $H''$ state, whereas the light triplet
is simply decoupled.

\subsection*{3. Solution of the $\mu$ problem}

 The present approach allows for a simple
solution of the $\mu$ problem through the introduction of a light
gauge singlet superfield $N$ \cite{nm}.
The VEV of $N$ induced after SUSY breaking
plays the role of an effective mass term  $\mu H \bar H$ in the
low-energy theory. The corresponding part of the superpotential
has the following form:
\begin{equation}
W_{\mu} = \lambda N10^2 + \lambda'N^3/3.
\end{equation}
This form is the most general under the $Z_2\otimes Z_3$
symmetry introduced above,
provided $N$ is invariant under $Z_2$, whereas it
transforms in the same way as $10$ under $Z_3$.
In order to guarantee the decoupling of the $N$ and $10$ from the
heavy GUT Higgs fields, we require that none of them transform under
$Z_3$.

Such a simple solution of the $\mu$ problem
is very difficult to implement in the standard cases with a large
D--T mass hierarchy, since the introduction of a light singlet
normally leads to a disastrous destabilization of the hierarchy
through the
well-known one-loop `tadpole' diagram \cite{tp}.
The same difficulty appears in different
versions of the `sliding singlet' scenario \cite{sp}. The source of
the trouble is that the light singlet couples to both
light ($H,\bar H$) and heavy
($T, \bar T$) states; because of this, its exchange immediately induces
masses and VEVs of the $H, \bar H$ of the order of the geometric
mean $\sim (M_{GUT}m_s)^{1/2}$, where $m_s$ is a SUSY-breaking scale
in the low-energy sector. This is a serious problem for any scenario
(with light singlets)
in which the D--T masses are split. In contrast, such a difficulty never
occurs in our case, since doublet and triplet are both light `by
definition' and $N$ does not couple to the heavy states. So the troublesome
`tadpole' is absent, allowing for a simple solution of the $\mu$ problem.

 It is worth pointing out that even in the standard GUTs with the heavy
triplet partner the $\mu$ problem
may be solved by some other mechanism. For example,
when embedded in the minimal
supergravity with the hidden sector SUSY breaking \cite{hs},
the $\mu$ problem is automatically solved in the `pseudo-Goldstone picture'
\cite {pp}; $\mu = m_{3/2}$ is induced by a shift of heavy
VEVs triggered by the SUSY breaking. In the other schemes, the solution
may be achieved by going beyond the minimal supergravity and
introducing the couplings of the $10$-plet with the hidden sector
fields in the non-minimal K\"ahler potential \cite {gm}
(although one may need some effort in order
to explain why the similar couplings are absent from the superpotential).
However, the solution with light singlet can work equally well even in
the schemes with much lower scale of SUSY breaking, where the
above supergravity solutions do not work.

\subsection*{4. SO(10) examples }
Now, let us turn
to the model building and produce two realistic
$SO(10)$ examples. The sectors of the theory that are responsible for the
proton stability, fermion masses and the $\mu$ problem we have
discussed above in a more or less model-independent way (apart from the
fermion-mass structure, of course, which we believe has to be addressed
in the frame of some specific flavour symmetry).
In fact what we need now is to take care of the Higgs
sector that breaks GUT symmetry and which does not participate
in  $W_{Yukawa}$ or $W_{\mu}$ due to $Z_2\times Z_3$-symmetry.
As we know, the only GUT
Higgs allowed to speak with $W_{Yukawa}$ is the
$45$-plet with the VEV (2). Below we will denote it as $A$.
The requirements that the GUT Higgs superpotential ($W_{GUT}$)
has to obey are the following:

 ($a$) $W_{GUT}$ should be most general under symmetries;

 ($b$) no `fine-tuning';

 ($c$) it should allow for the $G_w$-symmetric SUSY minimum in which
$A$ has a VEV $(2)$ and all particles except for one
pair of $L,\bar L$-type states + complete
$SU(5)$ multiplets + (possibly) some $G_W$-singlets,
have the GUT scale mass in order to keep the successful unification of
gauge couplings \cite{uc} intact.

We present below two models.

\subsubsection*{4.1 Model I}
$W_{GUT}$ includes the chiral superfields in the following
$SO(10)$ representations: $ S,X,Y \equiv$ singlets;
$\Sigma \equiv 54$-plet; $A,B,C,\Phi \equiv$ 45-plets;
$\chi, \bar {\chi}, \psi, \bar {\psi} \equiv 16,\overline{16}$-plets
and $F \equiv 10$-plet
(not to be confused with the $10$-plet in $W_{Yukawa}$). The superpotential
has the form:
\begin{eqnarray}
W_{GUT} &=& {\sigma \over 4}STr\Sigma^2 + {h \over 6} Tr\Sigma^3 +
{1 \over 4}Tr(a\Sigma + M_a + a'S)A^2 +
{1 \over 4}Tr(b\Sigma + M_b + b'S)B^2 \nonumber\\
&+& {1 \over 2}Tr(a''XA + b''YB)C + {g_c \over 2}\bar {\chi}C\chi +
g_f\chi F \chi + \bar {g_f}\bar {\chi} F\bar {\chi} +
g_{\Phi}\bar {\psi} \Phi \chi\nonumber\\
&+& \bar {g_{\Phi}}\bar {\chi} \Phi \psi +
g_a\bar {\psi}A \psi
+ \rho X Tr\Phi^2 + M^2S + {M' \over 2}S^2 + {\kappa \over 3}S^3
\end{eqnarray}
This form is strictly natural, since it is the most general compatible
with the $Z_4^A\otimes Z_2^B \otimes U(1)^C$ global symmetry under
which the chiral superfields  transform as follows:

 under $Z_4^A$
\begin{eqnarray}
 && (A,X,10) \rightarrow -(A,X,10)\nonumber\\
 && (\psi,\bar {\psi}) \rightarrow  i (\psi,\bar {\psi})\nonumber\\
 && \phi \rightarrow -i\phi
\end{eqnarray}
 under $Z_2^B$
\begin{equation}
(B,Y) \rightarrow -(B,Y)
\end{equation}
and under $U(1)^C$
\begin{eqnarray}
 (C,F) &\rightarrow& e^{i2\alpha} (C,F)\nonumber\\
 (\chi,\bar {\chi}) &\rightarrow& e^{-i\alpha}(\chi,\bar {\chi})\nonumber\\
 (X,Y) &\rightarrow& e^{-i2\alpha} (X,Y)\nonumber\\
 \Phi &\rightarrow& e^{i\alpha} \Phi
\end{eqnarray}
As the reader can observe, $Z_4^A$ acts as $Z_2$ on $A$ and $10$. This
is precisely the same $Z_2$ symmetry as was introduced in
section 2 and which forces $the 10$-plet to be
coupled with the matter superfields
only in combination with $A$. We assume that all mass scales in $W_{GUT}$
are $\sim M_{GUT}$ and all coupling constants are of the order of 1.

The standard procedure shows that the above superpotential admits
the following supersymmetric ($F$-flat and $D$-flat) minimum with
an unbroken $G_W$ symmetry:
\begin{eqnarray}
\Sigma &=& diag(2,2,2,2,2,2,-3,-3,-3,-3)\Sigma~~~~where~
\Sigma = {b'M_a - a'M_b \over 3ab' + 2ba'}\nonumber\\
A &=& diag[0,0,0,A,A]\otimes \epsilon\nonumber\\
B &=& diag[B,B,B,0,0]\otimes \epsilon\nonumber\\
\chi &=& \bar {\chi} = \chi|+,+,+,+,+\rangle~~
where~\chi^2 = -{a'' \over g_c} XA = -{b'' \over g_c} YB\nonumber\\
S &=& -{2bM_a + 3aM_b \over 3ab' + 2ba'}\nonumber\\
\psi &=& \bar {\psi} = F = \Phi = C = 0
\end{eqnarray}
According to the standard notations (e.g. see \cite{wz}) the
$SU(5)$ singlet component of $16$ is denoted
by $|+,+,+,+,+\rangle$, where each $`+'$
refers to an eigenvalue
of the respective Cartan subalgebra generator.
The two quantities $A$ and $B$ are determined
from the two equations:
\begin{eqnarray}
&& 10(S\sigma \Sigma - h\Sigma^2) - aA^2 + bB^2 =0\nonumber\\
&& 15\sigma\Sigma^2 + a'A^2 + {3 \over 2}b'B^2 + M^2 +M'S +
\kappa S^2 =0
\end{eqnarray}
Note that the absolute VEVs of the singlets $X$ and $Y$ are undetermined
in the SUSY limit, and only their ratio, ${X \over Y} = {b''B \over a''A}$,
is. It is not difficult to check that in the given vacuum $W_{GUT}$
delivers a pair of light doublets (with quantum numbers of $L, \bar L$)
from the $\psi, \bar {\psi}$ multiplets. This is because
$\psi, \bar {\psi}$ states get their masses only from the two sources:
through the VEV of $A$ and via mixing with the heavy $\Phi$ through $\chi,
\bar {\chi}$
VEV. Now, the $A$ VEV leaves all $SU(2)_L$-doublet states in
$\psi,\bar {\psi}$ massless. These are the states with quantum numbers
$Q, \bar Q$ and $L, \bar L$, which in the $SU(5)$ language belong to
$10,\bar {10}$ and $5, \bar 5$ representations respectively;
$10,\overline{10}$ components are mixed with similar fragments of $\Phi$
through the $SU(5)$ singlet VEVs of $\chi, \bar {\chi}$ and become heavy.
In contrast, $5, \bar 5$ states
cannot do so, since they have no partners in the
$45$-plet $\Phi$. Thus, $L, \bar L$ states are massless. All other
$G_W$ non-singlet states from $W_{GUT}$ have a GUT scale mass. If we
recall now that in the $W_{Yukawa}$ sector we already had one light
triplet pair ($T, \bar T$) on top of the MSSM particle content, it will
be clear that new light states form a complete $SU(5)$-multiplets
($5,\bar 5$) and the unification of couplings is thus unaltered.

\subsubsection*{4.2 Model II}

In this version the $SO(10)$ content of $W_{GUT}$ is the same as in
the Model I except for the
fact that we exclude the $10$-plet $F$, the $45$-plet $\Phi$,
and the two singlets $X$ and $Y$ from the
theory. The superpotential becomes:
\begin{eqnarray}
W_{GUT} &=& {\sigma \over 4}STr\Sigma^2 + {h \over 6} Tr\Sigma^3 +
{1 \over 4}Tr(a\Sigma + M_a + a'S)A^2 +
{1 \over 4}Tr(b\Sigma + M_b + b'S)B^2\nonumber\\
&+& {1 \over 4}Tr(c\Sigma + M_c + c'S)C^2 +
g_a\bar {\psi} A \psi + {1 \over 2} g_b\bar {\chi} B \chi +
\bar {\chi}(M'' + \gamma S)\chi\nonumber\\
&+& M^2S + {M' \over 2}S^2 + {\kappa \over 3}S^3
\end{eqnarray}
Again, this form is natural in the strong sense, as it is most general
under $Z_4^A\otimes Z_2^C$ symmetry, which acts on the chiral superfields
in the following way:

 under $Z_4^A$ (as before)
\begin{equation}
(A,10) \rightarrow  - (A,10),~~~~~
(\psi,\bar {\psi}) \rightarrow  i (\psi,\bar {\psi})
\end{equation}
 under $Z_2^C$
\begin{equation}
 C \rightarrow  - C.
\end{equation}
All other superfields are invariant under the given symmetries.
Again, by the straightforward solution of the standard $F$-flatness and
$D$-flatness conditions we can find the following $G_W$-preserving
supersymmetric vacuum
\begin{eqnarray}
\Sigma &=& diag(2,2,2,2,2,2,-3,-3,-3,-3)\Sigma
{}~~~~where ~\Sigma = {c'M_a - a'M_c \over 3ac' + 2ca'}\nonumber\\
A &=& diag[0,0,0,A,A]\otimes \epsilon\nonumber\\
B &=& diag[B,B,B,B',B']\otimes \epsilon\nonumber\\
C &=& diag[C,C,C,0,0]\otimes \epsilon\nonumber\\
\chi &=& \bar {\chi} = \chi|+,+,+,+,+>\nonumber\\
\psi &=& \bar {\psi} = 0\nonumber\\
S &=& -{2cM_a + 3aM_c \over 3ac' + 2ca'}
\end{eqnarray}
The five remaining
quantities $A,B,B',C,$ and $\chi$ are determined from the five
equations:
\begin{eqnarray}
&& (2b\Sigma + M_b + b'S)B + g_b\chi^2 = 0\nonumber\\
&& (-3b\Sigma + M_b + b'S)B'+ g_b\chi^2 = 0\nonumber\\
&& \gamma S + M'' + {g_b \over 2}(3B + 2B') = 0\nonumber\\
&& 10(\sigma S\Sigma - h\Sigma^2) - aA^2 + b(B^2 - B^{'2}) + cC^2 =0\nonumber\\
&& 15\sigma\Sigma^2 + a'A^2 + b({3 \over 2} B^2 + B^{'2}) +
   {3 \over 2}c'C^2 + \gamma \chi^2 + M^2 +M'S + \kappa S^2 =0
\end{eqnarray}
Again, in the above vacuum there is a set of $G_W$ non-singlet massless
states delivered by $W_{GUT}$. First of all, there are  $Q,\bar Q,
L,\bar L$ states from $\psi, \bar {\psi}$, which are zero eigenstates
of $T_R^3$ generator and cannot get masses from the $A$ VEV.
On top of this, there are the pseudo-Goldstone-type massless
(in the SUSY limit) states resulting from the continuous degeneracy of the
given vacuum. This degeneracy occurs because
one can continuously rotate the $A$
and $B$ VEVs by an arbitrary $independent$ global $SO(4)$ transformation
and/or $C$ and $B$ by an $independent$ global $SO(6)$ transformation
without violating any of the conditions $F = 0$ or $D = 0$.
This happens because
$A,B,$ and $C$ do not communicate directly with one another in the
superpotential, but only through $\Sigma$ to whom all $45$-plets are
coupled bilinearly; since the $\Sigma$ VEV is invariant under
$G_{LR}$, the vacuum automatically gets a larger degeneracy under
$SO(6)_C\otimes SO(6)_B$ and  $SO(4)_A\otimes SO(4)_B$ global
transformations. Thus, there are pseudo-Goldstone modes with quantum
numbers of the $SO(6)/SU(3)\otimes U(1)$ and $SU(2)_R/U(1)_R$
generators ($u_c \bar {u_c}$  and $e_c \bar {e_c}$
states) which are
not eaten up by the gauge superfields. Thus again, as in Model I we
end up with the complete $SU(5)$ multiplets beyond the MSSM particle
content, but now these new light states effectively compose a 4th
vector-like family $5 +\bar 5, 10 + \overline{10}$.
This preserves the successful
unification of the gauge couplings.

\subsection*{5. conclusions}

 We have presented an alternative approach to the D--T splitting, problem
which in contrast to the standard schemes $does~not$ require the
heavy coloured triplet Higgs. The crucial point is that, independently
from the triplet mass, the proton decay can be extremely suppressed
for group theoretical reasons if the quark and lepton masses are
induced from the high-dimensional operators of the form (1).
In this case, the light coloured triplet $automatically$ gets decoupled
after the GUT symmetry breaking.
We have shown how the desired operators can be
naturally (purely due to a symmetry and the field content)
induced after integrating out some heavy states at $M_{GUT}$.
Two serious problems of the standard approach: colour-Higgsino-mediated
proton decay and the $\mu$ problem, can receive a natural solution. The
first one is automatic: both Higgsino- and Higgs-mediated proton
decays are suppressed by the same rate $(\sim M_W/M_{GUT})^2$ so that the
proton is stable (practically). The second problem can be easily solved
by introducing a
light gauge singlet superfield without causing the standard
`light singlet' problem.

 Another model-independent consequence is the existence of some decoupled
long-lived particles in the low-energy theory. These necessarily include
a coloured triplet Higgs pair (and at least one extra doublet pair which
automatically preserves the successful unification of couplings).
These new particles can be subject of an experimental search.
The extremely small Yukawa coupling constant
(suppressed at least by a factor $\sim {M_W \over M_{GUT}}$ with respect to
the ordinary doublet) makes the lightest member of the supermultiplets
$T,\bar T$ long-lived enough to appear stable in
the detector so that the colour singlet bound states, which they form
with ordinary quarks, should behave as heavy (with mass $\sim M_w$)
stable hadrons (or mesons). In this respect their phenomenology is very
similar to the one of the coloured pseudo-Goldstone states discussed in
\cite{lt},
although their origin is very different.

   Finally, we have presented two $SO(10)$ examples which naturally
accommodate the above scenario. Both have superpotentials that are most
general under symmetries and do not suffer from any fine-tuning problem.

\end{document}